\RequirePackage{fix-cm}

\documentclass[referee,sn-mathphys,Numbered]{sn-jnl}

\usepackage[T1]{fontenc}
\usepackage{lmodern}

\usepackage{graphicx}
\graphicspath{{img/}}
\DeclareGraphicsExtensions{.pdf} 

\usepackage{booktabs}
\usepackage{multirow}
\usepackage[table,dvipsnames]{xcolor}

\usepackage{amsmath,amssymb,amsfonts}
\usepackage{mathtools}
\usepackage{amsthm}
\usepackage[mathscr]{euscript}
\usepackage{bm}
\usepackage{braket}
\usepackage{empheq}

\usepackage[version=4]{mhchem}
\usepackage{siunitx}

\usepackage{caption}
\usepackage{subcaption}
\DeclareCaptionSubType*[arabic]{figure}

\usepackage{algorithm}
\usepackage{algorithmicx}
\usepackage{algpseudocode}
\usepackage{listings}

\usepackage[title]{appendix}
\usepackage[inline]{enumitem}
\usepackage{comment}
\usepackage[nohyperlinks,nolist]{acronym}
\usepackage{multicol}
\usepackage{manyfoot}
\usepackage{textcomp}
\usepackage{soul}
\usepackage{etoolbox}

\usepackage{todonotes}

\renewcommand{\thefigure}{\arabic{figure}}

\newlength{\dch}
\newlength{\sch}

\newcommand{\scaling}{\phi}

\newcommand{\major}{\cellcolor{blue!100}}
\newcommand{\minor}{\cellcolor{blue!25}}

\newcommand{\vampyr}{\texttt{VAMPyR}\cite{battistella_2023_10290360}}
\newcommand{\remrchem}{\texttt{ReMRChem}\cite{ReMRChem}}
\newcommand{\mrchem}{\texttt{MRChem}\cite{Wind_2023}}
\newcommand{\mrcpp}{\texttt{mrcpp}\cite{bast_2023_7967323}}
\newcommand{\grasp}{\texttt{GRASP}\cite{grasp_ref}}

\begin{acronym}
\acro{ABGV}{Alpert, Beylkin, Ginez and Vozovoi}
\acro{ae}{almost everywhere}
\acro{AO}{atomic orbital}
\acro{API}{Application Programmer Interface}
\acro{AUS}{Advanced User Support}
\acro{BEM}{Boundary Element Method}
\acro{BO}{Born-Oppenheimer}
\acro{BS}{B-Spline}
\acro{BVP}{boundary value problem}
\acro{CBS}{complete basis set}
\acro{CC}{Coupled Cluster}
\acro{CoE}{Centre of Excellence}
\acro{CTCC}{Centre for Theoretical and Computational Chemistry}
\acro{DC}{dielectric continuum}
\acro{DD}{domain decomposition}
\acro{DCHF}{Dirac-Coulomb Hartree-Fock}
\acro{DFT}{density functional theory}
\acro{DHF}{Dirac--Hartree--Fock}
\acro{DIIS}{Direct Inversion in the Iterative Subspace}
\acro{DKH}{Douglas-Kroll-Hess}
\acro{ECP}{effective core potential}
\acro{EFP}{effective fragment potential}
\acro{EU}{European Union}
\acro{FMM}{fast multipole method}
\acro{GGA}{generalized gradient approximation}
\acro{GPE}{Generalized Poisson Equation}
\acro{GTO}{Gaussian Type Orbital}
\acro{HF}{Hartree-Fock}
\acro{HPC}{high-performance computing}
\acro{Hylleraas}[HC]{Hylleraas Centre for Quantum Molecular Sciences}
\acro{IEF}{Integral Equation Formalism}
\acro{IEFPCM}{Integral Equation Formalism \ac{PCM}}
\acro{IGLO}{individual gauge for localized orbitals}
\acro{KAIN}{Krylov-accelerated inexact Newton}
\acro{KB}{kinetic balance}
\acro{KS}{Kohn-Sham}
\acro{KTRS}{Kramers' Time Reversal Symmetry}
\acro{LAO}{London atomic orbital}
\acro{LAPW}{linearized augmented plane wave}
\acro{LDA}{local density approximation}
\acro{MAD}{mean absolute deviation}
\acro{maxAD}{maximum absolute deviation}
\acro{MCSCF}{multiconfiguration self consistent field}
\acro{MM}{molecular mechanics}
\acro{MO}{molecular orbital}
\acro{MPA}{multiphoton absorption}
\acro{MRA}{multiresolution analysis}
\acro{MSDD}{Minnesota Solvent Descriptor Database}
\acro{MW}{multiwavelet}
\acro{NAO}{numerical atomic orbital}
\acro{NeIC}{nordic e-infrastructure collaboration}
\acro{NMR}{nuclear magnetic resonance}
\acro{NP}{nanoparticle}
\acro{NS}{non-standard}
\acro{OLED}{organic light emitting diode}
\acro{PAW}{projector augmented wave}
\acro{PBC}{Periodic Boundary Condition}
\acro{PCM}{polarizable continuum model}
\acro{PDE}{partial differential equation}
\acro{PW}{plane wave}
\acro{QC}{quantum chemistry}
\acro{QM}{quantum mechanics}
\acro{QM/MM}{quantum mechanics/molecular mechanics}
\acro{RCN}{Research Council of Norway}
\acro{RDM}{1-body reduced density matrix}
\acro{RI}{resolution of the identity}
\acro{RKB}{restricted kinetic balance}
\acro{RMSD}{root mean square deviation}
\acro{SAPT}{symmetry-adapted perturbation theory}
\acro{SC}{semiconductor}
\acro{SCF}{self-consistent field}
\acro{SCRF}{self-consistent reaction field}
\acro{SERS}{surface-enhanced raman scattering}
\acro{SI}{Supporting Information}
\acro{STSM}{short-term scientific mission}
\acro{TRS}{time-reversal symmetry}
\acro{WP}{Work Package}
\acro{WPREL}[WP1]{Work Package 1}
\acro{WPROP}[WP2]{Work Package 2}
\acro{WPAPP}[WP3]{Work Package 3}
\acro{X2C}{exact two-component}
\acro{ZORA}{zero-order relativistic approximation}
\end{acronym}

\begin{document}

\title{Four-component relativistic calculations in a multiwavelet basis with improved convergence}

\author[1]{Jacopo Masotti}
\author[1,2]{Roberto Di Remigio Eik\aa s}
\author*[3]{Christian Tantardini}\email{christiantantardini@ymail.com}
\author*[1]{Luca Frediani} \email{luca.frediani@uit.no}

\affil[1]{Hylleraas center, Department of Chemistry, UiT The Arctic University of Norway, PO Box 6050 Langnes, N-9037 Troms\o, Norway}
\affil[2]{Algorithmiq Ltd., Kanavakatu 3C, FI-00160, Helsinki, Finland}
\affil[3]{Center for Integrative Petroleum Research, King Fahd University of Petroleum and Minerals, Dhahran 31261, Kingdom of Saudi Arabia}

\abstract{
We revive an approach to solve the Dirac equation originally proposed by Kutzelnigg which makes use of the squared Dirac operator $\hat{\mathfrak{D}}^{2}$. This approach holds the promise to avoid the negative energy solution because the negative energy spectrum is now ``folded" on the positive energy side and at the same time provides a convex equation, which is amenable to a minimization process and increased precision in the final result. 
The $\hat{\mathfrak{D}}^{2}$ yields an equation similar to the non-relativistic one, yet in a four-component framework, where Multiwavelet tools and algorithms developed for the non-relativistic case can be employed with minor modifications. 
On the other hand, the use of Multiwavelets is here essential to achieve the full potential of the approach.
We implemented and validated this approach for one- and two-electron systems with increasing nuclear charge. Numerical tests were performed to gauge the actual precision of the approach with respect to either analytical reference values when possible or numerical results obtained with the \textit{GRASP} code otherwise.
}

\keywords{multiwavelets, Dirac equation, squared-Dirac operator, Kutzelnigg, relativity}



\maketitle
\section{Introduction}
\label{sec:intro}

Relativistic effects are not a niche in molecular electronic structure: for heavy elements they can qualitatively change bonding, spectra, and response properties through scalar-relativistic contraction/expansion and spin--orbit coupling.
The four-component Dirac equation is the natural starting point for a first-principles description of electrons whenever high precision is required beyond the nonrelativistic Schr\"odinger equation.\cite{Dirac_1928,Pyykko_2012}

Despite its conceptual clarity, a direct variational treatment of the Dirac operator is delicate because its spectrum is \textit{unbounded} from below. 
In finite basis representations this can manifest as the well-known \textit{variational collapse}  and related pathologies when the discretization inadequately separates positive- and negative-energy states.\cite{Mark1980-zz,Wallmeier_1981,Mark_1982,Schwarz_1982,dyall2012relativistic,dyall2009relativistic,dyall2007relativistic,gomes2010relativistic,dyall2006relativistic,visscher1991kinetic,gianturco1971some,liu2010ideas,stanton1984kinetic,tatewaki2003prolapses}

A practical way to restore a well-posed variational framework is to replace the Dirac operator by a spectrally safer surrogate that preserves the physical eigenfunctions.
The squared Dirac operator, $\hat{\mathfrak{D}}^{2}$, is bounded from below and admits a standard Rayleigh--Ritz treatment while retaining the same eigenfunctions only with squared eigenvalues.\cite{Wallmeier_1981,Wallmeier1984-ta} 
This makes $\hat{\mathfrak{D}}^{2}$ particularly attractive when the numerical method is variational, and when one seeks systematic convergence rather than basis-set empiricism.

In parallel, adaptive \acp{MW}  bases provide a route to all-electron calculations with guaranteed, tunable precision and without conventional basis-set error.
They refine locally where the solution demands it as close to nuclei and in bonding regions as in asymptotic tails, while keeping a compact representation elsewhere.\cite{Harrison_2004,Anderson_2019}
Recent works have demonstrated that fully numerical four-component Dirac calculations for molecules are feasible in such adaptive \ac{MW} bases.\cite{Anderson_2019,tantardini2023future} 
What is still missing is an equally systematic \ac{MW} realization of $\hat{\mathfrak{D}}^{2}$, which would combine (i) the variational robustness of the squared formulation with (ii) the precision control and adaptivity of \ac{MRA}.

The manuscript is organized as follows. 
In Sec.~\ref{sec:theo}, we review the theoretical background of the four-component Dirac operator and its multiresolution formulation, including the adaptive MW basis and the integral-equation form of the Dirac--Fock problem. 
In Sec.~\ref{sec:dev}, we develop the squared-Dirac-operator formulation within the multiwavelet framework. 
There, the many-electron problem is treated at the instantaneous Dirac--Coulomb level, so that the mean-field potential contains the usual direct and exchange contributions, $\hat{J}$ and $\hat{K}$, together with the nuclear attraction term. 
The generalized potential entering the $\hat{\mathfrak{D}}^{2}$ formulation, the cross-expectation-value analysis, and the use of an artificial speed of light to amplify relativistic effects are also introduced in this section. 
Numerical results for one- and two-electron systems are reported in Sec.~\ref{sec:results}, where we assess stability, attainable precision, derivative-operator dependence, grid-placement effects, and computational performance. 
Finally, Sec.~\ref{sec:conclusions} summarizes the main conclusions and outlines the steps needed to move from the present proof-of-concept implementation toward a production-grade relativistic multiwavelet code.

\section{Theoretical Background}
\label{sec:theo}

Hereinafter, we will use Hartree atomic units, and for clarity, we will keep the electron mass $m$ explicit in the equations.
The free-space Dirac operator is given as:
\begin{equation}
\hat{h}_{D} = c \Vec{\alpha} \cdot \Vec{p} + \beta mc^{2}
\end{equation}
where $\Vec{p}$ collects the Cartesian components of the momentum operator:
\begin{equation}
\Vec{p} = -\mathrm{i}\begin{pmatrix}
\frac{\partial}{\partial x} \\
\frac{\partial}{\partial y} \\
\frac{\partial}{\partial z} 
\end{pmatrix}
\end{equation}
and $\Vec{\alpha}$ is a Cartesian vector whose components are $4\times 4$ anti-diagonal block matrices:
\begin{equation}
\alpha_{u} = 
\begin{pmatrix}
0 & \sigma_{u} \\
\sigma_{u} & 0
\end{pmatrix}
\end{equation}
We use indices $u, w \in \lbrace x, y, z \rbrace$ for Cartesian components. The $2\times 2$ matrices $\sigma_{u}$ appearing as anti-diagonal blocks are the Pauli matrices. The $4\times 4$ matrix $\beta$ is defined as:
\begin{equation}
    \beta = 
    \begin{pmatrix}
    I_{2} & 0 \\
    0     & -I_{2}
    \end{pmatrix}
\end{equation}

The eigenfunctions of the free-space Dirac operator are vectors with four complex components:
\begin{equation}
    \Phi
    = 
    \begin{pmatrix}
    \varphi^{1} \\ 
    \varphi^{2} \\
    \varphi^{3} \\
    \varphi^{4}
    \end{pmatrix},
\end{equation}
the upper (lower) two components of the orbital can equivalently be referred to as \emph{large} (\emph{small}) components and indexed accordingly as:
\begin{equation}
    \Phi
    = 
    \begin{pmatrix}
    \varphi^{L_{\alpha}} \\ 
    \varphi^{L_{\beta}} \\
    \varphi^{S_{\alpha}} \\
    \varphi^{S_{\beta}}
    \end{pmatrix}.
\end{equation}
The large and small components can be transformed into each other. Practically, the exact relationship is only applicable for the free-particle problem. In atomic and molecular applications, one initializes the small components by applying the \emph{kinetic balance} condition, i.e. the nonrelativistic limit of the exact large-small relationship, to the large components \cite{sun2011comparison}.

\subsection{Adaptive Multiwavelets Basis}

\Ac{MRA} is a numerical framework devised to solve integral and differential equations with arbitrary, predefined precision.\cite{Alpert_MRA}
Standard quantum chemical methods expand the unknown \acp{MO} into \emph{fixed} atom-centered bases and recast the solution of the \ac{HF} equations
as a matrix eigenvalue problem to determine the expansion coefficients~\cite{JensenBook}.
\ac{MRA}-based methods instead solve directly for the occupied \acp{MO}, through adaptive refinement of the expansion basis.
The mathematical framework enabling such adaptive algorithms relies on the construction of a sequence of orthogonal subspaces which is dense in the space of square-integrable functions for a user-defined simulation box.

As an example, consider the unit interval and a set of orthonormal $k$-th order polynomials, $\scaling_{i}(x)$, which we call \emph{scaling} functions. The basis can be refined by \emph{dyadic} subdivisions of the interval: at any subdivision \emph{scale}, $n$,
the original unit interval is subdivided into $2^{n}$ sub-intervals. Each interval supports a \emph{fixed} number of scaling functions, which are obtained
by dilation and translation of the same functions at scale $n=0$:
\begin{equation}
    \scaling^n_{il} = 2^{n/2} \scaling_i(2^n x - l),
\end{equation}
where $l=0,\ldots,2^{n-1}$ is the translation index, identifying functions in the sub-interval $[l/2^n, (l+1)/2^n]$.
This construction defines a \emph{telescopic} series of scaling spaces:
\begin{equation}
 V_{0}^{k} \subset V_{1}^{k} \subset ..... V_{n}^{k}\subset ....\subset L^{2}([0, 1]),
\end{equation}
The limit of this series is dense in $L^{2}([0, 1])$ and can thus represent \emph{any} $f \in L^{2}([0, 1])$. 
The orthogonal complement between two successive scaling spaces constitutes a wavelet space:
\begin{equation}
 V_{n}^{k} \oplus W_{n}^{k} = V_{n+1}^{k},\: \: \: W_{n}^{k} \perp V_{n}^{k},
\end{equation}

Several features of a \ac{MW} approach\cite{Alpert_1993} concur to efficient algorithms with tunable and controlled precision in quantum chemistry\cite{Harrison_2004, Frediani_2013}. The vanishing moments of the wavelet basis lead to compact (small) representation of functions\cite{Alpert_MRA}, disjoint support enables adaptivity which enhances compactness, the \ac{NS}-form of operators\cite{Beylkin101016jacha200708001} effectively decouples scales at the critical step of the application of convolution operators, the availability of separated representation of operators\cite{Beylkin_ACHA} alleviates the so-called ``curse of dimensionality".

In practice, this allows to control the error of a calculation within a range defined by a precision parameter $\epsilon$ which typically varies from $10^{-4}$ (labeled MW4)  to $10^{-8}$ (labeled MW8) in relative terms with respect to the norm of a target function.
Numerical efficiency is maximized when the order of the polynomial $k$ is chosen such that $k \simeq -\log_{10} \epsilon + 3$.

For a MW$n$ ($n= 4, 5, \ldots, 8$) calculation, convergence is considered achieved once the norm of the orbital update drops below $\epsilon = 10^{-n}$. For a variational approach this implies that the error in the energy should be $10^{-2n}$.

However, as we showed in our previous work on the full Breit Hamiltonian \cite{tantardini2023future}, requesting tighter precision for the Dirac equation, demands a higher polynomial order than what provided by default by our \ac{MW} library, \mrcpp{}. 

The interested reader is referred to the available literature for additional details about \ac{MRA} and corresponding details about how to achieve the requested precision in its practical realizations for quantum chemistry.\cite{Alpert_1993, Beylkin2005-kg, Frediani_2013, Harrison_2003, Anderson_2019}

A feature of a \ac{MW} representation and related orbital optimization is that the whole $L^2$ space is spanned, with the requested precision being the only limitation. As such this is the ideal approach to revive the original idea by Kutzelnigg and Wallmeier in \cite{Wallmeier_1981}, according to which, $\hat{\mathfrak{D}}^{2}$ yields more precise results than the standard $\hat{\mathfrak{D}}$. As they point out, this approach requires a nearly complete basis, to make use of a \ac{RI} technique. That was prohibitive with \ac{AO} bases at the time. Multiwavelets, on the other hand, guarantee completeness to within any numerical precision available and this problem is therefore solved. It is possible that today's \ac{GTO} bases would perform better, although the \ac{RI} might be plagued by basis-set overcompleteness. \acp{MW} instead, provide completeness by construction.

\subsection{Integral equation formulation for the Dirac operator}

The Dirac-Fock equations are the 4-component, relativistic counterpart of the HF equations:
\begin{equation}\label{eq:dirac-scf-0}
    \hat{\mathfrak{D}}\Phi_i 
    = 
    (\hat{h}_{D} + \hat{V}) \Phi_{i} = (c \Vec{\alpha}\cdot \Vec{p} + \beta m c^{2} + \hat{V})\Phi_i 
    = 
    \sum_j D_{ij}\Phi_j, 
\end{equation}
where $\hat{h}_{D} = c \Vec{\alpha}\cdot \Vec{p} + \beta m c^{2}$ is the free-particle Dirac operator and $\hat{V} = \hat{J} - \hat{K} + \hat{V}_{\mathrm{nuc}}$ is the \ac{SCF} one-body potential (implicitly multiplied by the identity matrix $\mathbb{I}_4$). The matrix $D_{ij} = \braket{\Phi_i | \hat{\mathfrak{D}} | \Phi_j}$ is the projection of the mean-field Hamiltonian $\hat{\mathfrak{D}}$ in the occupied subspace spanned by $N$ orthonormal spinors. These are a set of coupled first-order differential equations. Its solutions are 4-component spinors; each component is a complex function and the set of solutions describes both positive and negative energy states.
We rearrange Eq.~\eqref{eq:dirac-scf-0} as:
\begin{equation}\label{eq:dirac-scf-1}
    (\hat{h}_{D} - D_{ii}) \Phi_{i} = - \hat{V} \Phi_{i} + \sum_{j\neq i} D_{ij}\Phi_j.
\end{equation}

The nonrelativistic Schr\"odinger equation, and its mean-field equivalent, can be solved by iterative convolution with the Green's function for the shifted Laplacian \cite{Kalos1962-ok}. 
Following \citet{Blackledge2013-jh}, \citet{Anderson_2019} formulated a similar iterative convolution for Eq.~\eqref{eq:dirac-scf-1}. We revise here the main steps of the derivation. First, we introduce the identity on the left-hand side:
\begin{equation}
\begin{aligned}
(\hat{h}_{D} - D_{ii})\underbrace{(\hat{h}_{D} + D_{ii})(\hat{h}_{D} + D_{ii})^{-1}}_{\mathbb{I}_{4}} \Phi_{i}
&= \left[\hat{h}_{D}^{2} - D_{ii}^{2}\right](\hat{h}_{D} + D_{ii})^{-1} \Phi_{i} \\
 &=
 - \hat{V} \phi_{i} + \sum_{j\neq i} D_{ij}\Phi_j,
 \end{aligned}
\end{equation}
The term in square brackets, thanks to Dirac's identity $(\Vec{\sigma}\cdot \Vec{p})(\Vec{\sigma}\cdot \Vec{p}) = p^2$, is  diagonal in the four components:
\begin{equation}
\hat{h}_{D}^{2} - D_{ii}^{2} = \left[ - c^{2} \nabla^2 + m^{2}c^{4} - D_{ii}^{2} \right]\mathbb{I}_{4}.
\end{equation}
This yields:
\begin{equation}
 \left[-\nabla^2 + \frac{m^{2}c^{4} - D_{ii}^{2}}{c^{2}}\right]I_{4}(\hat{h}_{D} + D_{ii})^{-1} \Phi_{i}
 = 
 - \frac{1}{c^2}\hat{V} \Phi_{i} + \frac{1}{c^{2}}\sum_{j\neq i} D_{ij}\Phi_j,
\end{equation}
which can be formally inverted to give:\cite{Beylkin2021-sk}
\begin{equation}\label{eq:dirac-scf-convolution}
    \Phi_{i} = - \frac{1}{c^{2}} \left[
    (\hat{h}_{D} + D_{ii}) 
    G_i
    \right]
    \star
 \left[ \hat{V} \Phi_{i} -\sum_{j\neq i} D_{ij}\Phi_j \right],
\end{equation}
w
since we are seeking bound states, \emph{i.e.} $D_{ii} < mc^{2}$.
As noted by \citet{Anderson_2019}, one can use the properties of the convolution product to rearrange Eq.~\eqref{eq:dirac-scf-convolution} as:

\begin{subequations}
\begin{empheq}[left={\Phi_{i} = - \frac{1}{c^{2}} \empheqlbrace}]{align}
    & \left[ (\hat{h}_{D} + D_{ii}) G_i \right] \star \left[ \hat{V} \Phi_{i} -\sum_{j\neq i} D_{ij}\Phi_j \right] \label{eq:A-dirac-scf-convolution} \\
    & (\hat{h}_{D} + D_{ii}) \left[ G_i \star \left(\hat{V} \Phi_{i} -\sum_{j\neq i} D_{ij}\Phi_j \right) \right] \label{eq:B-dirac-scf-convolution} \\ 
    & G_i \star \left[ (\hat{h}_{D} + D_{ii}) \left(\hat{V} \Phi_{i} -\sum_{j\neq i} D_{ij}\Phi_j \right) \right] \label{eq:C-dirac-scf-convolution} 
\end{empheq}
\end{subequations}
See also Appendix~\ref{app:equivalence-integral-ops} for a detailed derivation of these equivalences.
As we show in Appendix~\ref{app:15a-explicit}, the iterative convolution of Eq.~\eqref{eq:A-dirac-scf-convolution} uses the inverse-distance-cube kernel, which also appears in the context of the relativistic two-electron gauge potential \cite{Saue2011-qg}. As shown by some of us in Ref.~\cite{tantardini2023future}, application of this kernel is computationally intensive and not numerically robust.
The second form, Eq.~\eqref{eq:B-dirac-scf-convolution}, is computationally more efficient: we apply the same convolution kernel used in 
the nonrelativistic case first, smoothing out any sharp features introduced by multiplication with the potential, and then we compute 
its derivatives \cite{Anderson2019-bx}.
The final form in Eq.~\eqref{eq:C-dirac-scf-convolution} requires application of the derivative operator on $V\phi_i$, which might have non-smooth features and thus introduce numerical errors, convergence slowdown or excessive memory demands.
 
\section{Developing of Squared Dirac Operator within Multiwavelets Field}
\label{sec:dev}

Following \citet{Wallmeier_1981}, the square of the Dirac operator is positive and bounded from below, while obviously sharing eigenfunctions and eigenvalues with the original operator:
\begin{equation}
\begin{aligned}
    \hat{\mathfrak{D}}^{2}\Phi_i 
    &= 
    \left\lbrace c^{2}p^{2}I_{4} + c [\Vec{\alpha}\cdot \Vec{p}, \hat{V}]_{+} + m^{2}c^{4} + mc^{2} [\beta, \hat{V}]_{+} + \hat{V}^{2}I_{4}\right\rbrace\Phi_i \\
    &=
    \sum_{j}(\hat{\mathfrak{D}}^{2})_{ij}\Phi_j.
\end{aligned}
\end{equation}
Shifting by $m^{2}c^{4}$ and rescaling by $2mc^{2}$ yields:
\begin{equation}\label{eq:dirac-squared-scf-0}
    \left\lbrace 
    \frac{p^{2}}{2m}I_{4} + \hat{\mathcal{V}}
    \right\rbrace\Phi_i
    =
    \frac{1}{2mc^2}
    \sum_{j}\left[(\hat{\mathfrak{D}}^{2})_{ij} - \delta_{ij}m^2c^4\right]\Phi_j,
\end{equation}
where, for brevity, we define the \emph{generalized} mean-field potential:
\begin{equation}\label{eq:generalized-potential}
   \hat{\mathcal{V}} = 
    \frac{[\beta, \hat{V}]_{+}}{2}
    + \frac{[\Vec{\alpha}\cdot \Vec{p}, \hat{V}]_{+}}{2mc}
    + \frac{\hat{V}^{2}}{2mc^2}I_{4}.
\end{equation}

At convergence, the positive-energy orbital energies of $\hat{\mathfrak{D}}$ are $\epsilon_i = -mc^2 + \sqrt{m^2c^4 + 2mc^2\omega_i}$, where $\omega_i$ are the eigenvalues of the shifted-and-scaled \ac{SCF} equations above.
Eq.~\eqref{eq:dirac-squared-scf-0} is close to the nonrelativistic \ac{HF} equations, despite the appearance of more complicated potential-like terms. Once again, it can be recast as an integral equation and solved by iterative convolution: 
\begin{equation}\label{eq:dirac-squared-scf-convolution}
    \Phi_i 
    =
    -2m
    G_{i}
    \star
    \left[
    \hat{\mathcal{V}}\Phi_i
    +
    \frac{1}{2mc^2}\sum_{j\neq i}(\hat{\mathfrak{D}}^{2})_{ij}\Phi_j
    \right],
\end{equation}
We choose to approximate the matrix elements of the $\hat{\mathfrak{D}}^{2}$ operator as:
\begin{equation}\label{eq:matrix-product}
(\hat{\mathfrak{D}}^{2})_{ij} \simeq \sum_k D_{ik}D_{kj},
\end{equation}
in the \emph{converged} occupied subspace and for a complete basis (\acp{MW} are to be regarded as a complete basis to within their predefined precision), this relation would be exact.\footnote{
The MRA construction only explicitly refines the occupied orbital space and the \ac{SCF} procedure iteratively block-diagonalizes the matrix representation of the Fock operator. 
Away from convergence, the off-diagonal blocks will be nonzero:
\begin{equation*}
\begin{pmatrix}
\mathbf{O} & \mathbf{A} \\
\mathbf{A}^{\dagger} & \mathbf{V} 
\end{pmatrix}.
\end{equation*}
Forming the square of the Fock operator as the product gives:
\begin{equation*}
\begin{pmatrix}
\mathbf{O}^{2} + \mathbf{A}\mathbf{A}^{\dagger} & \mathbf{O}\mathbf{A} + \mathbf{A}\mathbf{V} \\
\mathbf{A}^{\dagger}\mathbf{O} + \mathbf{V}\mathbf{A}^{\dagger} & \mathbf{V}^{2} + \mathbf{A}^{\dagger}\mathbf{A}
\end{pmatrix}
\end{equation*}
showing that only when $\mathbf{A} = 0$ (at convergence) the representation in the occupied basis of $\mathbf{F}^{2}$ will be equal to the product of the basis representation of $\mathbf{F}$ with itself.
}
The exponent $\mu_i$ in Eq.~\eqref{eq:helmholtz-exponent} is now:
\begin{equation}
    \mu_i 
    = 
    \sqrt{\frac{m^2c^4 - (D^{2})_{ii} }{c^2}} 
    \simeq
    \sqrt{\frac{m^2c^4 - \sum_k D_{ik}D_{ki} }{c^2}}. 
\end{equation}
For the positive-energy states of interest, it always holds that $\mu_i > 0$.

\subsection{The generalized mean-field potential}

The iterative solution of the integral equation~\eqref{eq:dirac-squared-scf-convolution} requires the repeated application of the generalized mean-field potential in Eq.~\eqref{eq:generalized-potential}.
For general operators in a spinor basis it holds:
\begin{equation}
[\beta, M]_{+} = 
2 
\begin{pmatrix} 
M^{L_{\alpha}L_{\alpha}} & M^{L_{\alpha}L_{\beta}} & 0 & 0\\
M^{L_{\beta}L_{\alpha}}  & M^{L_{\beta}L_{\beta}} & 0 & 0 \\
0      & 0 & -M^{S_{\alpha}S_{\alpha}} & -M^{S_{\alpha}S_{\beta}} \\
0      & 0 & -M^{S_{\beta}S_{\alpha}}  & -M^{S_{\beta}S_{\beta}} 
\end{pmatrix}.
\end{equation}
In a \ac{MW} framework, we work directly with the \ac{MO} basis: thus the potential operator $V$ is \emph{diagonal} in the spinor components and its anticommutator with $\beta$ is simply:
\begin{equation}
[\beta, \hat{V}]_{+} =  2 \beta \hat{V}.
\end{equation}

Application of $\hat{\mathcal{V}}$ involves the, potentially demanding, construction of the squared mean-field potential.
Formally, its application on a test spinor is:
\begin{equation}
   \hat{V}^{2}\Phi_k =  
   \hat{J}^2\Phi_k + \hat{K}^{2}\Phi_k + \hat{V}_{\mathrm{nuc}}^{2}\Phi_k
   +
   [\hat{J} - \hat{K}, \hat{V}_{\mathrm{nuc}}]_{+}\Phi_k
   - 
   [\hat{J}, \hat{K}]_{+}\Phi_k,
\end{equation}
however, since we are working in a first-quantization setting, these operators can be applied as:
\begin{equation}
   \hat{V}^{2}\Phi_k = \hat{V}\left(\hat{V}\Phi_k\right).
\end{equation}


For the one- and two-electron systems we consider in this work, the generalized potential is greatly simplified.
For one-electron systems, we have:
\begin{equation}\label{eq:one-generalized-potential}
   \hat{\mathcal{V}} = 
    \beta \hat{V}_{\mathrm{nuc}}
    + \frac{[\Vec{\alpha}\cdot \Vec{p}, \hat{V}_{\mathrm{nuc}}]_{+}}{2mc}
    + \frac{\hat{V}_{\mathrm{nuc}}^{2}}{2mc^2}I_{4},
\end{equation}

For closed-shell, two-electron systems, we can explicitly exploit Kramers' \ac{TRS} and remove the exchange terms:\cite{Kramers1930-hq}
\begin{equation}\label{eq:two-generalized-potential}
   \hat{\mathcal{V}} = 
    \beta (\hat{J} + \hat{V}_{\mathrm{nuc}})
    + \frac{[\Vec{\alpha}\cdot \Vec{p}, \hat{J} + \hat{V}_{\mathrm{nuc}}]_{+}}{2mc}
    + \frac{(\hat{J} + \hat{V}_{\mathrm{nuc}})^{2}}{2mc^2}I_{4},
\end{equation}

\subsection{Cross expectation values}
We can use two different approaches to converge our result toward the ground state: the Dirac-Fock integral equation \eqref{eq:C-dirac-scf-convolution}, or the Squared Dirac Operator integral equation \eqref{eq:dirac-squared-scf-convolution}. Both algorithms will effectively evolve the initial guess spinor function towards the ground state energy solutions, by propagating the component functions in the Hilbert space until the total update will be lower than the set threshold \cite{Beylkin2021-sk}. Once convergence is reached, we can evaluate the final ground state energy either by computing the expectation value of $\hat{\mathfrak{D}}$ or $\hat{\mathfrak{D}}^{2}$. 

To better discuss the methods and the results, we will denote by $\varepsilon(\hat{\mathcal{O}}_E, \Phi_{\hat{\mathcal{O}}_{SCF}})$ the orbital energy obtained by taking the expectation value of $\hat{\mathcal{O}}_E$ for the wavefunction $\Phi_{\hat{\mathcal{O}}_{SCF}}$, obtained by performing the SCF using the $\hat{\mathcal{O}}_{SCF}$ operator.
We will thus have four different possibilities:
\begin{equation}\label{eq:diagonal_terms_matrix}
    \begin{pmatrix}
        \varepsilon(\hat{{\mathfrak{D}}}, \Phi_{\hat{{\mathfrak{D}}}}) & 
        \varepsilon(\hat{{\mathfrak{D}}}^2, \Phi_{\hat{{\mathfrak{D}}}}) \\ 
        \varepsilon(\hat{{\mathfrak{D}}}, \Phi_{\hat{{\mathfrak{D}}}^2}) & 
        \varepsilon(\hat{{\mathfrak{D}}}^2, \Phi_{\hat{{\mathfrak{D}}}^2}) 
    \end{pmatrix}
\end{equation}
The diagonal elements represent results where both the wave function optimization and the energy expectation value are obtained with the same method. The off-diagonal terms represent results where the wave function optimization and the energy are calculated using different methods. This adds a further dimension of variables to investigate that could lead to more precise result by using less resources.

\subsection{Amplification of relativistic effects for lighter atoms}\label{sec:amplification}
It is  sometimes convenient to be able to run calculation with an \emph{artificial} speed of light. For higher values one expects to get non-relativistic results, whereas for lower values one can amplify relativistic effects. The latter is particularly useful here to understand if the observed precision (or lack thereof) should be attributed to the relativistic effect or to the nuclear charge. For example, assume we want to achieve the same relativistic correction for \ce{Ar} as one would observe for \ce{Hg}.
We can impose that the new speed of light, when inserted in the analytical formula for the 1-electron atom in the case of a $1s$ electron for Argon, would yield the same relative energy shift as in Mercury with the physically correct $c$. Given the analytical solution for the ground state of a single electron system for the Dirac equation in atomic units:
\begin{equation}
    E^R(c) = c^2  \sqrt{1-\frac{Z^2}{c^2}}- c^2, 
\end{equation},
we want to obtain the following:
\begin{equation}
    \frac{E^{NR}_{\ce{Ar}} - E^{R}_{\ce{Ar}}(c')}{E^{NR}_{\ce{Ar}}}= \frac{E^{NR}_{\ce{Hg}} - E^{R}_{\ce{Hg}}(c)}{E^{NR}_{\ce{Hg}}}
\end{equation}
This is verified if $c' = \frac{Z_{\ce{Ar}}}{Z_{\ce{Hg}}} \;c$.

\section{Results and Discussion}
\label{sec:results}

The goal of the numerical tests presented here is to asses the precision of $\hat{\mathfrak{D}}^{2}$ with respect to $\hat{\mathfrak{D}}$ in the MW framework. As a reference we have taken the results obtained with the \grasp{} code.
The tests will investigate the precision obtained for both methods, employed both in the \ac{SCF} optimization and to compute the orbital energy. This yields a total of four combination as outlined in Eq. \eqref{eq:diagonal_terms_matrix}. We also assessed  the effects of the choice of derivative operator (\ac{BS}~\cite{ANDERSON2019100033} \emph{vs.} \ac{ABGV}~\cite{Alpert_MRA} realizations) and grid choice (nuclear charge on a dyadic point or offset from it).

We have carried out a series of tests on 1- and 2- electron systems with increasing nuclear charge. For the 1-electron systems we have considered \ce{H}, \ce{He^+}, \ce{Ne^{9+}}, \ce{Ar^{17+}} and  \ce{Hg^{79+}}, whereas for 2-electron systems we have included \ce{He}, \ce{Ne^{8+}}, \ce{Ar^{16+}} and  \ce{Hg^{78+}}.

Except when specified otherwise, the speed of light was set to $c=137.0359895\; \textit{a.u.}$. For the two smallest nuclei (\ce{H} and \ce{He}) we made use of a simple point-like charge model.
For the \ac{MW} calculations the usual numerical, precision-dependent smoothing described in~\cite{Yanai2004} was employed, to avoid an excessive refinement of the grid when trying to represent the potential of the nucleus.
For heavier nuclei (\ce{Ne}, \ce{Ar}, \ce{Hg}) we employed a Fermi-Dirac charge distribution \cite{Visscher_1997} as this is the model available in the \grasp{} software as we have done in a previous study \cite{tantardini2023future}.
In Table~\ref{tab:Fermi-Dirac-parameters} 
we report the explicit parameters employed in the two codes. The difference is due to the conventions adopted. In \remrchem{}, the skin thickness is defined as described in~\cite{Visscher_1997}, with a constant $a=2.3$fm (then converted in a.u.), while the half charge radius varies from atom to atom. \grasp{} uses the same physical model but it employs a thickness parameter rescaled by the factor $4\ln{(3)}$, which appears explicitly in the original article~\cite{Visscher_1997}. This is why the $a$ parameter is different in the two inputs. We emphasize this technical point to ease reproducibility and because it required quite some effort to track down this subtlety when comparing our results to \grasp{}.

\begin{table}[h]
\centering
\caption{All the values are given and used in atomic units (Bohr radii). 
}\label{tab:Fermi-Dirac-parameters}
\begin{tabular}{c|SS|SS}
\hline
 & \multicolumn{2}{c|}{Grasp} & \multicolumn{2}{c}{VAMPyR} \\
nucleus & {$a$} & {$c$} & {$a$} & {$c$} \\
\hline
\ce{Ne} & {$9.89059\, 10^{-6}$} & {$5.589069\, 10^{-5}$} & {$4.34637\, 10^{-5}$} & {$5.589069\, 10^{-5}$} \\
\ce{Ar} & {$9.89059\, 10^{-6}$} & {$6.883946\, 10^{-5}$} & {$4.34637\, 10^{-5}$} & {$6.883946\, 10^{-5}$} \\
\ce{Hg} & {$9.89059\, 10^{-6}$} & {$1.2433249\, 10^{-4}$} & {$4.34637\, 10^{-5}$} & {$1.2433249\, 10^{-4}$} \\
\hline
\end{tabular}
\end{table}

All \ac{MW} results presented were obtained with our Python implementation in the \remrchem{} project, whose source code is freely available through GitHub.
Our implementation uses the \mrcpp{} library and its Python interface provided by the \vampyr{} package \cite{bjorgve2024}. This implementation can only deal with at most two-electron systems since it relies, as already stated, on Kramers' \ac{TRS} \cite{Kramers1930-hq,Wigner_1932} to reduce memory footprint and computational cost and can be considered a proof-of-concept implementation to validate the use of the squared Dirac operator within the \ac{MW} framework. 
This is the ideal starting point to consider inter-electronic interactions in the \ac{SCF} setting for the validation of the developed theory.
However, as it is known from the nonrelativistic \ac{MW} implementation in the \mrchem{} package (written in C++), increasing the number of electrons will not pose 
additional theoretical challenges, but only increase the required computational resources (memory in particular) as tested by our group and extensively discussed in Ref.~\citenum{Wind_2023}.

To assess the quality of the results, we performed a systematic set of calculations by varying four parameters in the relativistic code:
\begin{enumerate}[label=(\roman*),leftmargin=*,itemsep=2pt]
    \item the integral formulation used to optimize the eigenfunction, comparing the $\hat{\mathfrak{D}}$ and $\hat{\mathfrak{D}}^{2}$ operators;
    \item the requested precision, from $\delta=10^{-4}$ (MW4) to $\delta=10^{-8}$ (MW8);
    \item the derivative realization, comparing the ABGV and BS operators;
    \item the nuclear placement, either centered on a dyadic grid point or slightly displaced from it.
\end{enumerate}

The length of the box $L$ for the \ac{MRA} was dependent on the nuclear charge: $L=\lceil \frac{50}{Z}\rceil$, to make sure the electronic density is practically zero at the boundary. The threshold for \ac{SCF} convergence was set to $\epsilon = 10\,\delta$. The order of the polynomial was set to $k = 3-\log_{10}(\delta)$.
Convergence is achieved once the norm of the orbital update drops below $\epsilon$.

\subsection{One-electron calculations}
\label{sec:res_1}

For each of the selected nuclei, we have optimized the wave function and computed the final error in the energy compared to the reference provided by \grasp{}. The results have been collected in four different graphs. 
\begin{figure}
    \centering
    \includegraphics[width=\linewidth]{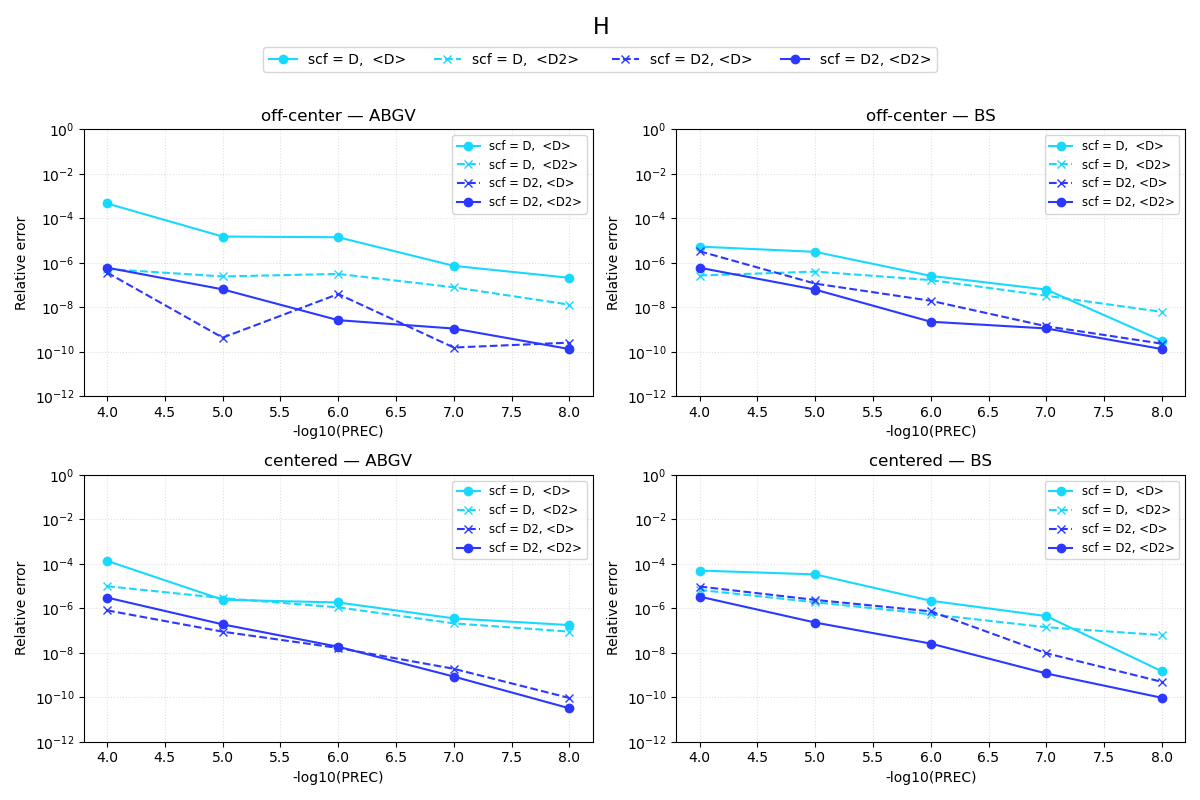}
    \caption{Relative error in the total energy of the H atom computed with \acp{MW}, with respect to the \grasp{} reference value. Each graph represents a choice of derivative operator (\ac{ABGV} or \ac{BS}) and atom placement (at a dyadic point or not). Each line in the graph represents a given choice of algorithm ($\Phi_{\hat{\mathfrak{D}}}$ or $\Phi_{\hat{\mathfrak{D}}^2}$) and expectation value ($\hat{\mathfrak{D}}$ or $\hat{\mathfrak{D}}^2$). Each point is the requested precision (from MW4 to MW8) of the calculation.}
    \label{fig:Rel_Err_1el_H}
\end{figure}
\begin{figure}
    \centering
    \includegraphics[width=\linewidth]{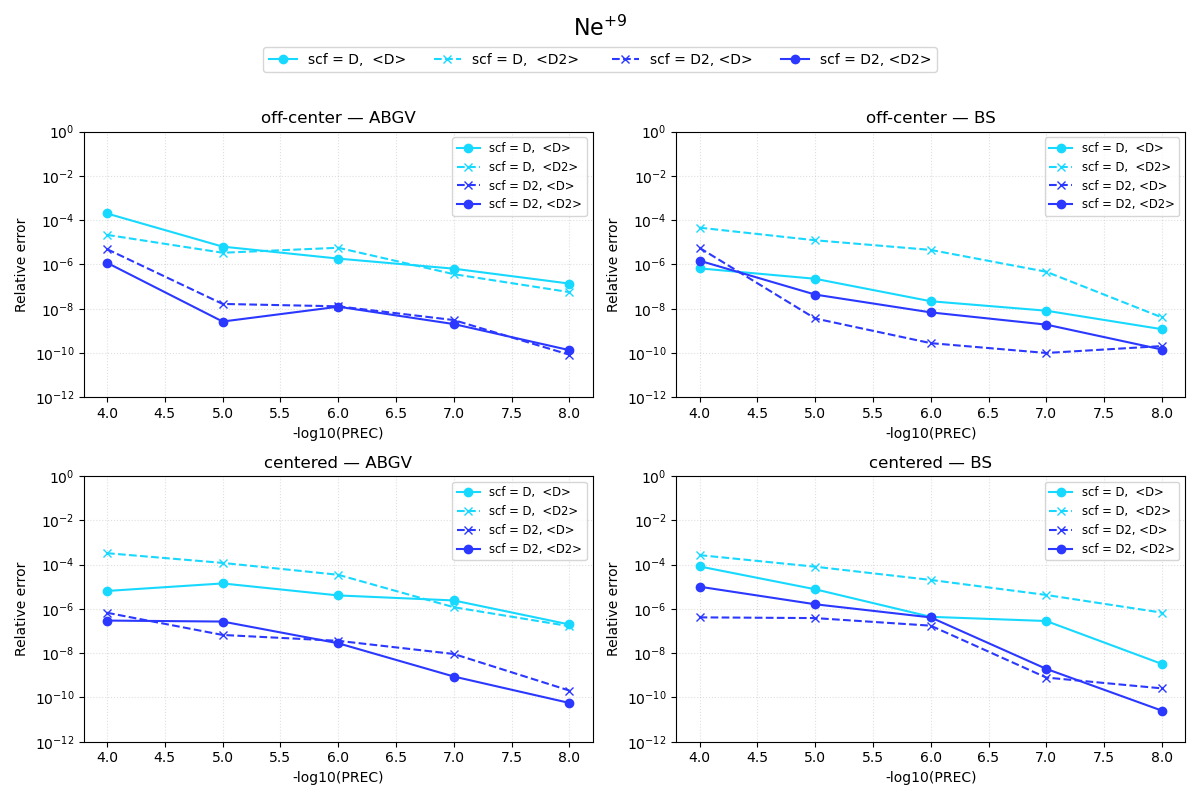}
    \caption{Relative error in the total energy of the \ce{Ne^{9+}} atom computed with \acp{MW}, with respect to the \grasp{} reference value. Each graph represents a choice of derivative operator (\ac{ABGV} or \ac{BS}) and atom placement (at a dyadic point or not). Each line in the graph represents a given choice of algorithm ($\Phi_{\hat{\mathfrak{D}}}$ or $\Phi_{\hat{\mathfrak{D}}^2}$) and expectation value ($\hat{\mathfrak{D}}$ or $\hat{\mathfrak{D}}^2$). Each point is the requested precision (from MW4 to MW8) of the calculation.}
    \label{fig:Rel_Err_1el_Ne}
\end{figure}
Each graph corresponds to one specific choice of derivative operator (ABGV or BS) and nucleus collocation (centered or off-centered with respect to the dyadic points of the grid). Each line in the graph corresponds to a specific pair of (a) optimization algorithm ($\Phi_{\hat{\mathfrak{D}}}$ in light blue and $\Phi_{\hat{\mathfrak{D}}^2}$ in dark blue) and (b) computation of the expectation value ($\hat{\mathfrak{D}}$ and $\hat{\mathfrak{D}}^2$).  Solid lines correspond to \emph{diagonal} setups in Eq.~\ref{eq:diagonal_terms_matrix}, whereas dashed lines are the \emph{off-diagonal} ones. For each line in the graph, different points correspond to a different requested precision from MW4 ($\delta=10^{-4}$) to MW8 ($\delta=10^{-8}$).

The results for \ce{H} and \ce{Ne^{9+}} are reported in \ref{fig:Rel_Err_1el_H} and \ref{fig:Rel_Err_1el_Ne}, all the other results are collected in the \ac{SI}.

The results show good agreement with the reference results given by \grasp{}: by tightening the precision parameter $\delta$, the difference with respect to \grasp{} narrows down as expected. We note however that by performing the \ac{SCF} optimization with the $\hat{\mathfrak{D}}^2$ operator, one generally achieves higher precision than the corresponding ones with the 
${\mathfrak{D}}$ operator, or in the worst case of similar precision. With reference to the symbolic matrix of methods in Eq.~\eqref{eq:diagonal_terms_matrix} the $\varepsilon(\hat{{\mathfrak{D}}}^2, \Phi_{\hat{{\mathfrak{D}}}^2})$ combination has shown to be consistently more precise than the corresponding $\varepsilon(\hat{{\mathfrak{D}}}, \Phi_{\hat{{\mathfrak{D}}}})$. 
When high precision is demanded, $\hat{\mathfrak{D}}^2$ calculations are able to reach errors of about $10^{-9}-10^{-10}$ consistently, whereas $\hat{\mathfrak{D}}$ results are often around two orders of magnitude less precise. In some cases the gap can reach four orders of magnitude. Off-diagonal combinations give a slightly mixed picture and don't seem to offer a clear advantage. 

Let us define the error of a given calculation as $\Delta \varepsilon(\hat{\mathcal{O}}_E = |\varepsilon(\hat{\mathcal{O}}_E, \Phi_{\hat{\mathcal{O}}_{SCF}})- \varepsilon_{ref}|$, using the corresponding \grasp{} result as a reference.

For what concerns the diagonal entries of \eqref{eq:diagonal_terms_matrix} and their respective errors, we observe that:
\begin{equation}
    \Delta\varepsilon(\hat{{\mathfrak{D}}}^2, \Phi_{\hat{{\mathfrak{D}}}^2}) \le  \Delta\varepsilon(\hat{{\mathfrak{D}}}, \Phi_{\hat{{\mathfrak{D}}}})
\end{equation} 
Since no clear pattern emerged for the off-diagonal terms, we'll only comment the results regarding the diagonal ones.

The choice of the derivative (\ac{ABGV} vs \ac{BS}) does not seem to affect the $\hat{\mathfrak{D}}^2$ calculations significantly. This is consistent with the fact that the only place where a derivative operator is applied in the $\hat{\mathfrak{D}}^2$ approach is in the anticommutator term in Equation~\ref{eq:generalized-potential}, prior to the convolution with the Helmholtz propagator. For the algorithm using the $\hat{\mathfrak{D}}$ operator, the derivative is instead applied \emph{after} the convolution, hence the BS version of the derivative leads to smoother functions in turns increasing precision as can be seen from Fig.~\ref{fig:Rel_Err_1el_H} and Fig.~\ref{fig:Rel_Err_1el_Ne}, in which the relative error (with respect to the \grasp{} reference) attained at MW8 is smaller compared to the same conditions using the \ac{ABGV} derivative. However, for tight thresholds (MW7-MW8), the choice of derivative has limited effect on the final precision for the $\hat{\mathfrak{D}}^2$ algorithm.
Putting the nucleus in a dyadic or a non-dyadic point, showed no clear effect. For this reason the results of charging this particular variable are inconclusive.

In summary, the $\hat{\mathfrak{D}}^2$ based algorithm generally showed an improvement in the precision obtained, with gains between 1 and 4 orders of magnitude. In a few cases, the results attained by using the two algorithms yielded roughly the same precision. 

\subsection{Two-electron calculations}
\label{sec:res_2}

\begin{figure}
    \centering
    \includegraphics[width=\linewidth]{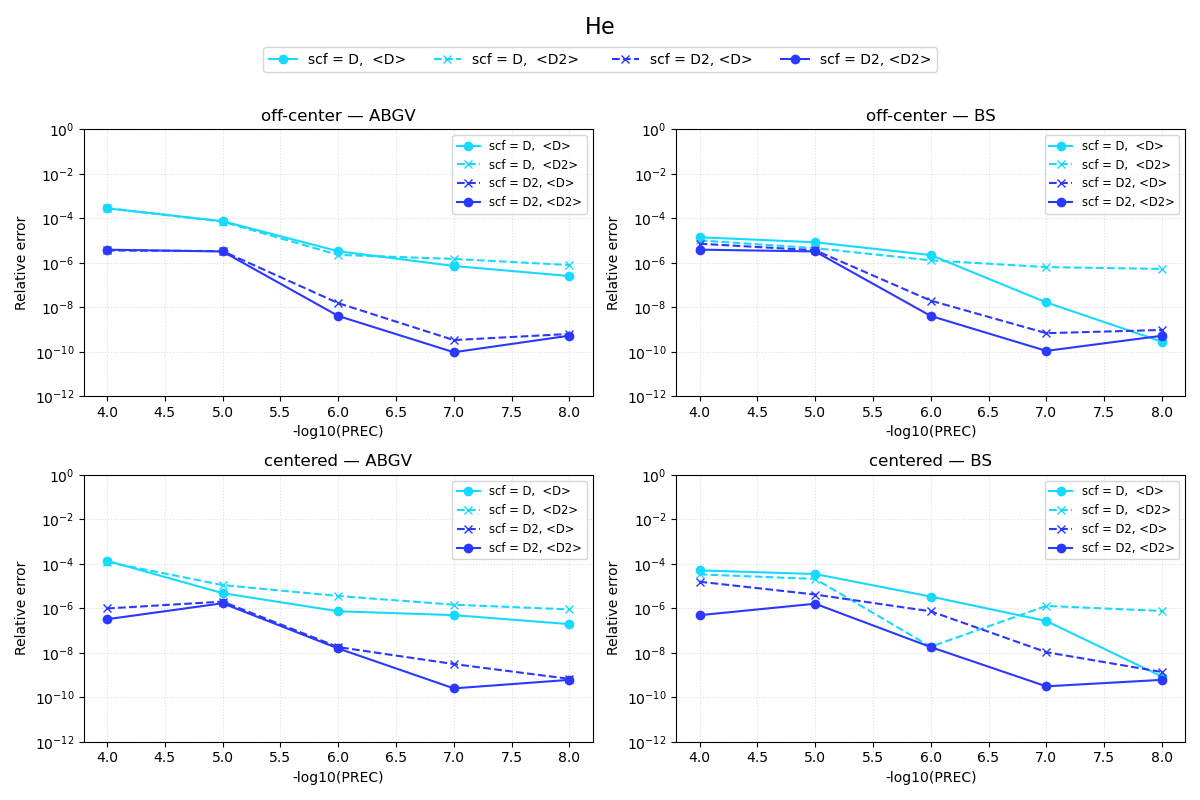}
    \caption{Relative error in the total energy of the He atom computed with \acp{MW}, with respect to the \grasp{} reference value. Each graph represents a choice of derivative operator (\ac{ABGV} or \ac{BS}) and atom placement (at a dyadic point or not). Each line in the graph represents a given choice of algorithm ($\Phi_{\hat{\mathfrak{D}}}$ or $\Phi_{\hat{\mathfrak{D}}^2}$) and expectation value ($\hat{\mathfrak{D}}$ or $\hat{\mathfrak{D}}^2$). Each point is the requested precision (from MW4 to MW8) of the calculation.}
    \label{fig:Rel_Err_2el_He}
\end{figure}

\begin{figure}
    \centering
    \includegraphics[width=\linewidth]{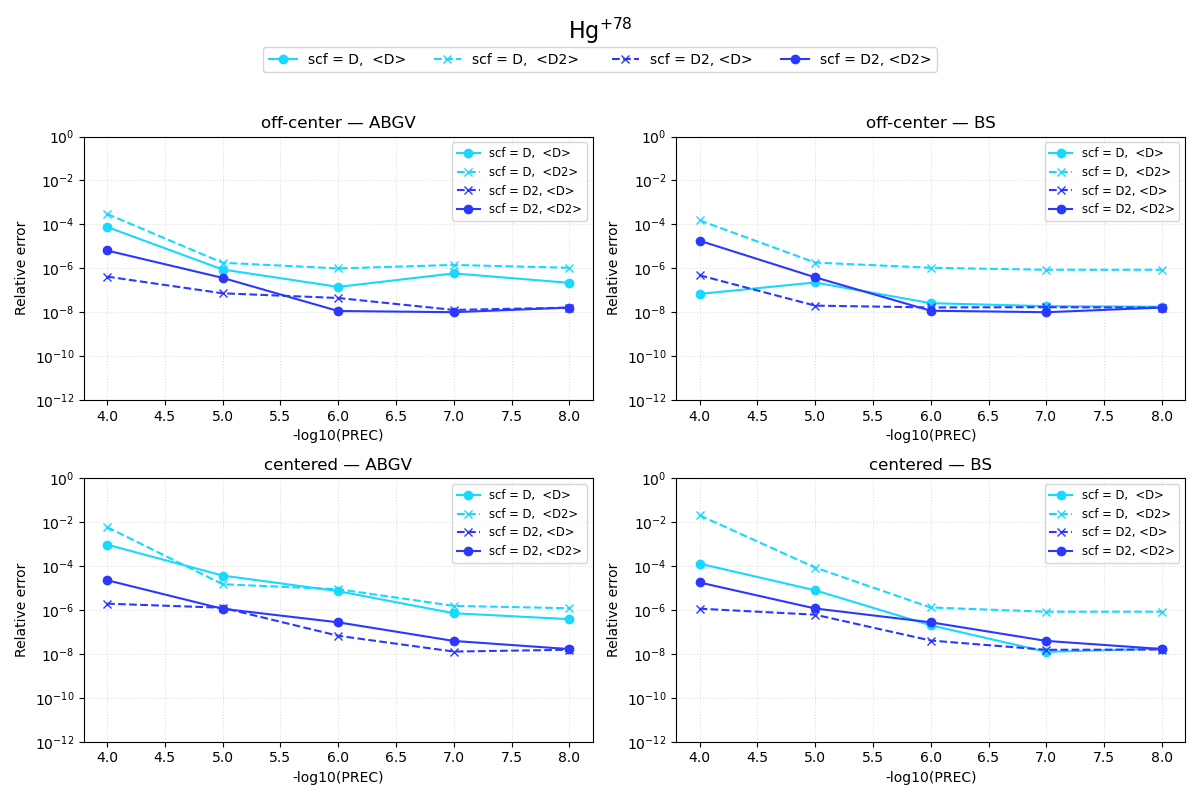}
    \caption{Relative error in the total energy of the Hg$^{+78}$ atom computed with \acp{MW}, with respect to the \grasp{} reference value. Each graph represents a choice of derivative operator (\ac{ABGV} or \ac{BS}) and atom placement (at a dyadic point or not). Each line in the graph represents a given choice of algorithm ($\Phi_{\hat{\mathfrak{D}}}$ or $\Phi_{\hat{\mathfrak{D}}^2}$) and expectation value ($\hat{\mathfrak{D}}$ or $\hat{\mathfrak{D}}^2$). Each point is the requested precision (from MW4 to MW8) of the calculation.}
    \label{fig:Rel_Err_2el_Hg}
\end{figure}

Two-electron calculations were carried out for the same nuclei, except \ce{H^-}, which is not amenable to our approach, using the bound-state Helmholtz propagator\footnote{Positive orbital energies lead to the complex-valued Helmholtz propagator which is currently not implemented.}. 
The results for He and \ce{Hg^{78+}} are reported in \ref{fig:Rel_Err_2el_He} and \ref{fig:Rel_Err_2el_Hg}, all the other results are collected in the supporting information.

We employed the instantaneous Dirac-Coulomb Hamiltonian, neglecting retardation effects of the Breit Hamiltonian and we exploited \ac{KTRS}. With these limitations, adding a second electron does not require the computation of a second spinor explicitly, and the electron-electron interaction requires only the inclusion of the direct term $J$ for a single electron and no exchange $K$: by virtue of \ac{KTRS}, the exchange term is in this case identical to the direct term, in analogy with a closed-shell non relativistic system.
As for the one-electron case, the ($\Phi_{\hat{\mathfrak{D}}^2}$, $\hat{\mathfrak{D}}^2$) combination is generally more precise than the corresponding ($\Phi_{\hat{\mathfrak{D}}}$, $\hat{\mathfrak{D}}$) one, with the exception of mercury as shown in figure \ref{fig:Rel_Err_2el_Hg} where the two approaches yield the same precision in the result. Moreover, the precision gain is far greater for lighter atoms, where the ($\Phi_{\hat{\mathfrak{D}}^2}$, $\hat{\mathfrak{D}}^2$) results are up to four orders of magnitude more precise than the ($\Phi_{\hat{\mathfrak{D}}}$, $\hat{\mathfrak{D}}$) ones. For heavier nuclei, such as \ce{Hg}, the precision gain observed by converging the wave function with $\hat{\mathfrak{D}}^2$ was generally less impressive, in the range of a single order of magnitude, or less.

For two-electron systems and heavier nuclei, it has also been necessary to apply a dampening of each iteration to achieve convergence: instead of using the result of the propagator equations \eqref{eq:dirac-scf-convolution} and \eqref{eq:dirac-squared-scf-convolution}, at each iteration we have taken a linear combination of the wave functions before and after the propagation. 

To understand the observed behavior for two-electron systems with heavier nuclei, we performed a series of calculations on \ce{Ar} with an artificial light speed $c' = 30.83309764$ as described in Sec.~\ref{sec:amplification}
These calculations required the same dampening strategy as for \ce{Hg} but provided the same precision obtained with \ce{Ar} and the standard speed of light. We conclude therefore that the deeper refinement needed do describe heavier nuclei is the limiting factor in the precision attained, rather than the choice of method ($\hat{\mathfrak{D}}$ vs $\hat{\mathfrak{D}}^2$). On the other hand, the magnitude of the relativistic correction affects the convergence of the calculation.

By comparing the relative error in the case of off-center BS calculations for Hg$^{+78}$ and Ar$^{+16}$ with the modified speed of light, we can observe that while in the first case there is a bottleneck in the precision attained, for the latter we don't see such a limitation and the error stately decreases as expected. We can therefore conclude that the bottleneck can be attributed to the absolute nuclear charge and not to the relativistic effect magnitude.
\begin{figure}
    \centering
    \includegraphics[width=\linewidth]{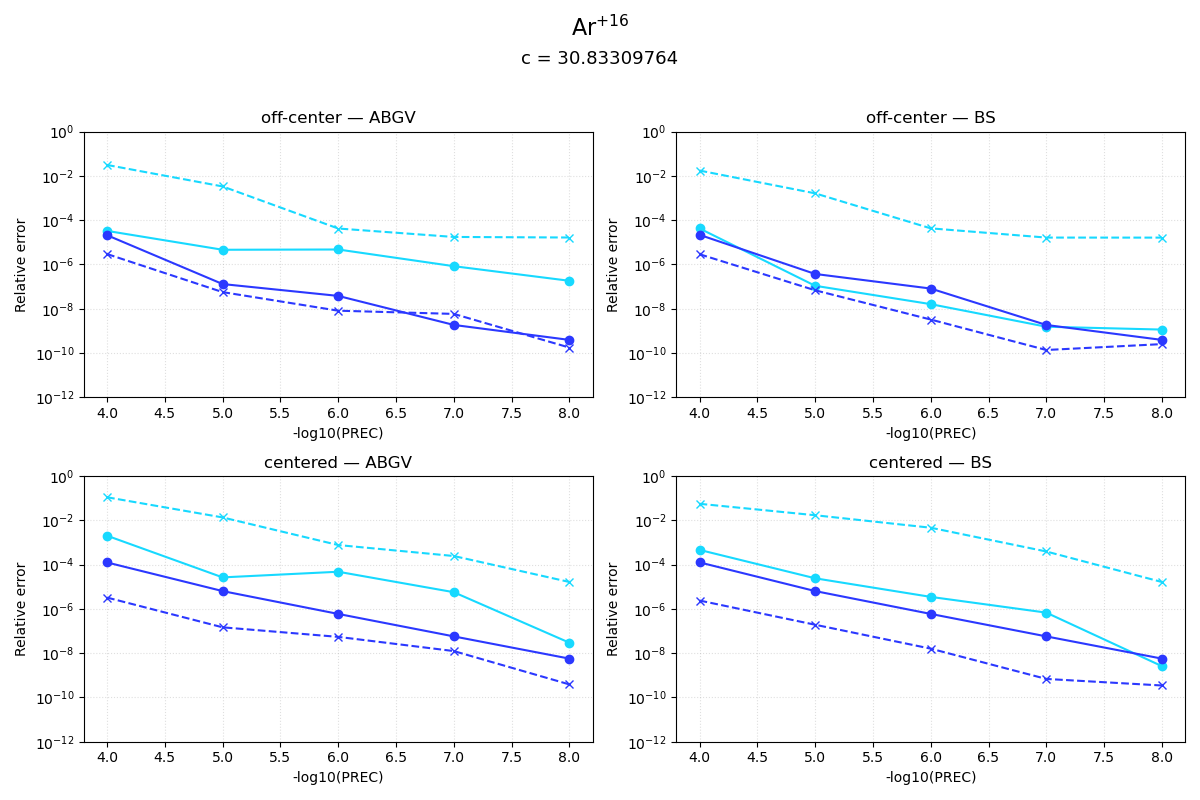}
    \caption{Relative error in the total energy of the \ce{Ar^{+16}} atom computed with \acp{MW}, with respect to the \grasp{} reference value and using a modified speed of light $c' = 30.83309764$ in order to achieve the same order of magnitude in the relativistic correction found in the Hg atom. Each graph represents a choice of derivative operator (\ac{ABGV} or \ac{BS}) and atom placement (at a dyadic point or not). Each line in the graph represents a given choice of algorithm ($\Phi_{\hat{\mathfrak{D}}}$ or $\Phi_{\hat{\mathfrak{D}}^2}$) and expectation value ($\hat{\mathfrak{D}}$ or $\hat{\mathfrak{D}}^2$). Each point is the requested precision (from MW4 to MW8) of the calculation.}
    \label{fig:placeholder}
\end{figure}

To investigate this further, we attempted to run some calculations using a Fermi-Dirac nuclear distribution for an artificial \ce{Hg} nucleus with a radius 10 and 100 times the regular one to achieve a smoother potential, requiring a less aggressive refinement in the vicinity of the nucleus. However, none of these calculations converged.

\subsection{Performance considerations}
\label{sec:res_3}

Although the $\hat{\mathfrak{D}}^2$-based algorithm is promising to achieve precise results, it requires a larger amount of memory, and each iteration is slower than the corresponding $\hat{\mathfrak{D}}$-based algorithm. This can be traced back to the need to compute more terms in the potential. Moreover, computing functions like $\hat{V}^2\Phi_i$, is numerically demanding because it requires a finer refinement of the grid close to the nucleus, further increasing the memory footprint and the time required for each iteration. This is especially evident for heavier nuclei where this is also the primary factor limiting the precision obtained. As noted above, two-electron systems also require dampening to reach convergence.

We underline that this is only a prototype implementation. The tradeoff between precision and performance will guide the design choices for a production-grade implementation. In such an implementation, it will moreover be possible to push precision requirements to achieve better results for heavier nuclei. Finally, the need for dampening will be superseded by off-the-shelf accelerators such as \ac{DIIS}~\cite{Pulay1980-xh} or \ac{KAIN}~\cite{Harrison101002jcc10108}.

A further possible performance gain for the $\hat{\mathfrak{D}}^2$-based algorithm, especially when working with larger molecular systems, would be to apply the different parts of the generalized potential in Eq.~\ref{eq:generalized-potential}
with a threshold scaled by their respective order in $c$.

We have considered two different realizations of the derivative operator in this work. The first one is the original version proposed by Alpert and coworkers\cite{Alpert_MRA}, whereas the second is a newer version based on BS\cite{Anderson2019-bx}. Derivative operators are challenging to implement with MW beacuse they are not band-limited and they can only be defined in a weak sense leading to a degree of arbitrariness. Such arbitrariness, can however be exploited to choose the best realization depending on the application\cite{Anderson2019-bx}. In this case, however, the choice of a specific type has yields only minor differences in terms of precision.

A more targeted visualization of the data can be found in the supporting information, showcasing the differences in precision obtained by the derivative type for any given type of algorithm discussed.
Comparing these figures for different atoms, we can conclude that when the expectation value of the energy was calculated using the $\hat{\mathfrak{D}}^2$ Hamiltonian, the results are less sensitive to the derivative choice compared to $\hat{\mathfrak{D}}$.

\section{Conclusions}
\label{sec:conclusions}

We have developed and implemented the squared Dirac operator, $\hat{\mathfrak{D}}^{2}$, in a \ac{MW} framework, following the original formulation  by Kutzelnigg and Wallmeier\cite{Kutzelnigg_1984,Wallmeier_1981}. The main motivation was to obtain a variationally well-posed relativistic mean-field scheme.  By squaring the Dirac operator, the negative-energy continuum is folded onto the positive-energy branch, resulting in an operator that is bounded from below. 
This transformation enables genuine energy minimization and systematic numerical convergence.
However, realizing these advantages in practice requires a numerical representation that does not introduce uncontrolled basis-set errors.
A \ac{MW} basis is crucial for this strategy. Unlike conventional atom-centered basis sets, \acp{MW} provide a near-complete representation within a prescribed tolerance, which is essential when matrix elements of $\hat{\mathfrak{D}}^{2}$ are expressed as products (see Eq.~\eqref{eq:matrix-product})

Furthermore, the squared formulation leads to an integral equation that closely resembles the non-relativistic Helmholtz-propagator strategy, allowing established multiresolution algorithms to be reused with limited modifications while retaining a four-component description.

Numerical tests on one- and two-electron atoms across increasing nuclear charge validate the implementation and show that the $(\hat{\mathfrak{D}}^{2},\Phi_{\hat{\mathfrak{D}}^{2}})$ scheme consistently outperforms $(\hat{\mathfrak{D}},\Phi_{\hat{\mathfrak{D}}})$ at fixed precision. 
At MW7--MW8, the $\hat{\mathfrak{D}}^{2}$-based SCF reaches $\sim 10^{-9}$--$10^{-10}$ absolute errors in the hardest one-electron cases, while $\hat{\mathfrak{D}}$-based SCF typically saturates one to two orders of magnitude earlier.
Cross variants show no systematic benefit, indicating the main gain comes from using $\hat{\mathfrak{D}}^{2}$ in the wavefunction optimization.

In this implementation, $\hat{\mathfrak{D}}^{2}$ is less sensitive to the derivative representation (ABGV vs.\ BS), and no systematic dependence on nuclear placement relative to the grid is observed, especially for tight thresholds (MW8). For two-electron systems and heavier nuclei, stable convergence required damped orbital updates.

The key drawback of our implementation is only computational, because the generalized potential $\hat{\mathcal{V}}$ introduces extra terms with notably repeated application of $V$ (effectively $V^{2}\Phi$), which increases near-nucleus refinement, memory, and wall time.
This was seen specifically for heavy nuclei where refinement already limits attainable precision.

Future work will address how to  to migrate from developing version to a production-grade parallel implementation with tighter thresholding, add robust SCF acceleration to remove ad hoc dampening, it will also extend the formulation to general many-electron molecules and relativistic Hamiltonians beyond the Dirac--Coulomb picture.

\backmatter

\section*{Data Availability}
All data supporting the findings of this study are provided within the article and the Supporting Information. The code used to generate the results and figures is publicly available at the \remrchem{} github repository.

\section*{Acknowledgments}
We would like to thank Prof.~Dr.~Trond Saue from the CNRS/Université de Toulouse, France for fruitful discussions on the topic.
We acknowledge support from the Research Council of Norway through its Centres of Excellence scheme (262695), through the FRIPRO grant ReMRChem (324590),  and from NOTUR -- The Norwegian Metacenter for Computational Science through grant of computer time (nn14654k).

\section*{Declarations}

We use the CRediT taxonomy of contributor roles.\cite{Allen2014-rd,Brand2015-qc}
The ``Investigation'' role also includes the ``Methodology'', ``Software'', and ``Validation'' roles.
The ``Analysis'' role also includes the ``Formal analysis'' and ``Visualization''
roles. The ``Funding acquisition'' role also includes the ``Resources'' role.
We visualize contributor roles in the following authorship attribution matrix,

\begin{table}[ht]
\caption{Levels of contribution: \textcolor{blue!100}{major}, \textcolor{blue!25}{minor}.}
\label{tbl:contribs}
\begin{tabular}{lcccc}
\hline\hline
                              &  JM    & CT     & RDRE   & LF     \\ \hline
    Conceptualization         &        & \major & \major & \major \\ 
    Investigation             & \major & \major & \minor & \minor \\ 
    Data curation             & \major &\major  & \minor & \minor \\ 
    Supervision               &        &\major  &        & \major \\ 
    Writing -- original draft & \major & \major & \major & \minor \\
    Writing -- revisions      & \minor & \major & \major & \major \\
    Funding acquisition       &        &        &        & \major \\ 
    Project administration    &        &  &        & \major \\ \hline\hline
\end{tabular}
\end{table}

\bibliography{sn-bibliography}

\clearpage
\begin{appendices}

\section{Equivalence of integral operator iterations}\label{app:equivalence-integral-ops}

The equivalence of the integral operator iterations in Eqs.~\eqref{eq:A-dirac-scf-convolution},~\eqref{eq:B-dirac-scf-convolution}, and~\eqref{eq:C-dirac-scf-convolution} rests 
on the relation between differentiation and convolution products:
\begin{equation}\label{eq:convolution-derivative}
(f \star g)^\prime = f \star g^\prime.
\end{equation}

We use square brackets to delimit the operands in the convolution. Expanding Eq.~\eqref{eq:A-dirac-scf-convolution}:
\begin{equation}
\begin{aligned}
\Phi_{i}  
=
&- \frac{1}{c^{2}} \left[ c (\Vec{\alpha} \cdot \Vec{p}) G_i \right] \star \left[ V \Phi_{i} - \sum_{j\neq i} D_{ij}\Phi_{j}\right] \\
&- \frac{1}{c^{2}} \beta mc^2 \left[ G_i \right] \star \left[ V \Phi_{i} - \sum_{j\neq i} D_{ij}\Phi_{j}\right] \\
&- \frac{D_{ii}}{c^{2}} \left[ G_i \right] \star \left[ V \Phi_{i} - \sum_{j\neq i} D_{ij}\Phi_{j}\right], 
\end{aligned}
\end{equation}
where both the scalar and $\beta mc^2$ terms have been moved outside of the convolution, \emph{i.e.} outside of the 
square brackets, the latter since $\beta I_4 = I_4 \beta$.
The first term in the convolution contains the derivative of $G_i$, thus we can use property~\eqref{eq:convolution-derivative} to move it outside of the convolution:
\begin{equation}
\begin{aligned}
\Phi_{i}  
=
&- \frac{1}{c^{2}} c (\Vec{\alpha} \cdot \Vec{p})  \left[ G_i \right] \star \left[ V \Phi_{i} - \sum_{j\neq i} D_{ij}\Phi_{j}\right] \\
&- \frac{1}{c^{2}} \beta mc^2 \left[ G_i \right] \star \left[ V \Phi_{i} - \sum_{j\neq i} D_{ij}\Phi_{j}\right] \\
&- \frac{D_{ii}}{c^{2}} \left[ G_i \right] \star \left[ V \Phi_{i} - \sum_{j\neq i} D_{ij}\Phi_{j}\right], 
\end{aligned}
\end{equation}
obtaining Eq.~\eqref{eq:B-dirac-scf-convolution}, after rearrangement.
From this form, we can use property~\eqref{eq:convolution-derivative} once more to move the derivative application on the $V\phi_i$ convolution partner,
\begin{equation}
\begin{aligned}
\Phi_{i}  
=
&- \frac{1}{c^{2}} \left[ G_i \right] \star \left[c (\Vec{\alpha} \cdot \Vec{p}) \left(V \Phi_{i} - \sum_{j\neq i} D_{ij}\Phi_{j}\right)\right] \\
&- \frac{1}{c^{2}} \beta mc^2 \left[ G_i \right] \star \left[ V \Phi_{i} - \sum_{j\neq i} D_{ij}\Phi_{j}\right] \\
&- \frac{D_{ii}}{c^{2}} \left[ G_i \right] \star \left[ V \Phi_{i} - \sum_{j\neq i} D_{ij}\Phi_{j}\right], 
\end{aligned}
\end{equation}
Eq.~\eqref{eq:C-dirac-scf-convolution} follows, since the corresponding rearrangement for the additional two terms in the convolution is trivial.

\section{Explicit form of the convolution kernel in Eq.~(16a)}
\label{app:15a-explicit}

Whereas the iterative integral operator application in Eqs.~\eqref{eq:B-dirac-scf-convolution} and~\eqref{eq:C-dirac-scf-convolution}
uses the (diagonal) matrix form of the nonrelativistic bound-state Helmholtz kernel, the iteration in Eq.~\eqref{eq:A-dirac-scf-convolution}
requires the use of a new convolution kernel: $c (\Vec{\alpha} \cdot \Vec{p}) G_i$.
This is a $4\times 4$ matrix convolution kernel:
\begin{equation}
c (\Vec{\alpha} \cdot \Vec{p}) G_i
=
c
\begin{pmatrix}
0 & 0 & \partial_z g_i & (\partial_x -\mathrm{i}\partial_y)g_i \\
0 & 0 & (\partial_x +\mathrm{i}\partial_y)g_i & -\partial_z g_i \\
\partial_z g_i & (\partial_x -\mathrm{i}\partial_y)g_i  & 0 & 0 \\
(\partial_x +\mathrm{i}\partial_y)g_i & -\partial_z g_i & 0 & 0
\end{pmatrix}
\end{equation}
with $g_i = \frac{1}{4\pi} \frac{\exp{(-\mu_i |\Vec{r} - \Vec{r}^\prime|)}}{|\Vec{r} - \Vec{r}^\prime|}$.
\newcommand{\derivBSH}[1]{
\left[
\left(\mu_i + \frac{1}{|\vec{r} - \vec{r}^\prime|}\right)
\frac{1}{|\vec{r} - \vec{r}^\prime|^2}
\right] 
(r_{#1} - r_{#1}^\prime)\exp{(-\mu_i |\vec{r} - \vec{r}^\prime|)}
}
The derivatives of the bound-state Helmholtz kernel are:
\begin{equation}
\begin{aligned}
\partial_u g_i 
&=
\frac{1}{4\pi} 
\left(
\mu_i \frac{(r_{u} - r_{u}^\prime)\exp{(-\mu_i |\vec{r} - \vec{r}^\prime|)}}{|\vec{r} - \vec{r}^\prime|^2}
+
\frac{(r_{u} - r_{u}^\prime)\exp{(-\mu_i |\vec{r} - \vec{r}^\prime|)}}{|\vec{r} - \vec{r}^\prime|^3}
\right) \\
&=
\frac{1}{4\pi} 
\derivBSH{u}
,\quad u \in \lbrace x, y, z \rbrace,
\end{aligned}
\end{equation}
leading to a $4\times 4$, anisotropic convolution kernel: 
\begin{equation}
\begin{aligned}
c (\Vec{\alpha} \cdot \Vec{p}) G_i
&=
\frac{c}{4\pi} 
\left[
\left(\mu_i + \frac{1}{|\vec{r} - \vec{r}^\prime|}\right)
\frac{1}{|\vec{r} - \vec{r}^\prime|^2}
\right] 
\exp{(-\mu_i |\vec{r} - \vec{r}^\prime|)} 
\times \\ &
\begin{pmatrix}
0 & 0 & (r_{z} - r_{z}^\prime) & (r_{x} - \mathrm{i}r_{y}) - (r_{x}^\prime - \mathrm{i}r_{y}^\prime) \\
0 & 0 & (r_{x} + \mathrm{i}r_{y}) - (r_{x}^\prime + \mathrm{i}r_{y}^\prime) & -(r_{z} - r_{z}^\prime) \\
 (r_{z} - r_{z}^\prime) & (r_{x} - \mathrm{i}r_{y}) - (r_{x}^\prime - \mathrm{i}r_{y}^\prime) & 0 & 0 \\
 (r_{x} + \mathrm{i}r_{y}) - (r_{x}^\prime + \mathrm{i}r_{y}^\prime) & -(r_{z} - r_{z}^\prime) & 0 & 0
\end{pmatrix}
\end{aligned}
\end{equation}

\end{appendices}

\end{document}